# Realizing spatiotemporal effective media for acoustic metamaterials

Xinhua Wen[1*], Xinghong Zhu[1*], Hong Wei Wu[1,2,3] and Jensen Li[1]

[1]Department of Physics, The Hong Kong University of Science and Technology, Clear Water Bay, Hong Kong, China
[2] School of Mechanics and Photoelectric Physics, Anhui University of Science and Technology, Huainan 232001, China
[3] National Laboratory of Solid State Microstructures, Nanjing University, Nanjing 210093, China

* authors with equal contributions

**Abstract**

The effective medium representation is fundamental in providing a performance-to-design approach for many devices based on metamaterials. While there are recent works in extending the effective medium concept into the temporal domain, a direct implementation is still missing. Here, we construct an acoustic metamaterial dynamically switching between two different configurations with a time-varying convolution kernel, which can now incorporate both frequency dispersion of metamaterials and temporal modulation. We establish the effective medium formula in temporally averaging the compressibilities, densities and even Willis coupling parameters of the two configurations. A phase disorder between the modulation of different atoms is found negligible on the effective medium. Our realization enables a high-level description of metamaterials in the spatiotemporal domain, making many recent proposals, such as magnet-free non-reciprocity, broadband slow-light and Fresnel drag using spatiotemporal metamaterials possible for implementations in future.

Metamaterials have demonstrated extraordinary capabilities in wave control for different classical waves in the past two decades [1-6]. While these materials are often composed of subwavelength structural units, an effective medium description of metamaterials is fundamental to provide a high-level description [7-10]. It further allows a performance-to-design approach for different optical and acoustic devices requiring an inhomogeneous profile of effective media, including invisibility cloaks, field rotators and metasurfaces [4, 11-13]. To further extend the capability of metamaterials, one direction is to make the material parameters being inhomogeneous in the temporal domain as well. In this case, either the geometric parameters of individual atoms, e.g. the volume of a Helmholtz resonator of an acoustic metamaterial with an external mechanical driver [14,15], or the resonating properties of the atoms with feedback circuits [16,17], can be tuned in real time. These initial implementations, together with many works associated to the exotic physics enabled by these time-varying parameters, have been opening up a new area, now being called spacetime (or spatiotemporal) metamaterials [18-21]. The breaking of the continuous translational symmetry in time provides an extra dimension in designing metamaterials, and allows us to explore magnet-free non-reciprocity [22-25], inverse prism [26], efficient frequency conversion[27-29], time-varying vortex beam [30], broadband slow-light [31], Floquet topological insulator and pump [32,33], etc.

Here, we are interested in the concept of effective medium, which is fundamental for describing metamaterials. For conventional metamaterials without time-varying parameters, termed as "static" hereafter, we require the wavelength being at least 5 times larger than the lattice constant [7,34] in order to have a valid effective medium description. A direct analogy of such concept in the temporal domain requires a modulation frequency at least a few times faster than the signal frequency [35,36]. Similarly, for a spatiotemporal effective medium, we need both criteria to be satisfied. However, the direct analogy of the effective medium concept in the temporal domain is yet to be implemented. This can be challenging, especially for electromagnetic (even for microwave) metamaterials, requesting both fine spatial control and fast temporal control. On the other hand, in the case of acoustic metamaterials, for the air-borne sound in the kHz regime in the current work, such a real-time tunability is feasible by adopting a digital feedback using a microcontroller [37,38], in which all digital calculations can be finished within one sampling period. The advantage of the digital feedback approach hinges on the huge flexibility in getting tailor-made resonating response and this flexibility can be extended to the time-varying case to achieve large changes in modulation amplitudes or resonating parameters. Moreover, various concepts in digital signal processing such as digital filtering and convolution techniques can be used as additional tools in molding the constitutive parameters with reconfigurable capability [37-39]. It allows such a digital feedback implementation in acoustics becomes a testbed for various proposed phenomena of spacetime metamaterials in the coming years. Here, we implement such a platform with fast modulations of the resonating frequency spectra of the constitutive parameters, such as compressibility (reciprocal of bulk modulus). With both the modulation period and lattice constant being much smaller than the signal period and signal wavelength respectively, we demonstrate an acoustic metamaterial in the spatiotemporal effective medium regime, giving a direct evidence that the effective medium formula is to temporally averaging the compressibilities and densities. Such a discussion can also be extended into the Willis coupling terms and random media with random modulation phases or random duty cycles of modulation for different atoms in the metamaterial.

Our implementation of spacetime metamaterial, for airborne sound waves running in a one-dimension waveguide, is schematically illustrated in Fig.1(a). The metamaterial in the middle section consists of several digitally virtualized resonating atoms, previously used for generating resonating compressibility, density and Willis coupling parameters with or without material gain in the static regime [37]. The inset of Fig. 1(a) shows a photograph of three virtualized atoms used in the current work. Each virtualized atom, labeled by an index $i$, has a speaker $S_i$ and a microphone $D_i$, which are interconnected by an external microcontroller. The microcontroller digitally samples the signal at the microphone and performs a convolution. To generate a time-varying resonating response (e.g. Lorentzian line-shape) at each atom, we choose to implement a time-varying convolution kernel with amplitude modulation, generally

written as

$$Y_i(t, t') = a(t)g \sin(\omega_0 t' + \theta_i) \exp(-\gamma_i t') \Theta(t') \quad (1)$$

so that the result of a time-varying convolution $S_i(t) = \int Y_i(t, t') D_i(t - t') dt'$ is feedback to the speaker in generating a resonating scattering response at each atom. For simplicity, we have written the convolution as if it is done in the continuous time domain as illustration without the digital representation. $\omega_0$ and $\gamma_i$ are the resonating (radial) frequency and linewidth respectively. $\theta_i$ is a line-shape tunning parameter. $\Theta$ is the Heaviside step function as the microcontroller can only implement a causal response. $g$ is a reference resonance strength with multiplication factor $a(t)$ in representing the temporal modulation of the resonance. When $a(t)$ is independent of $t$, the time-varying convolution kernel returns to a static one.

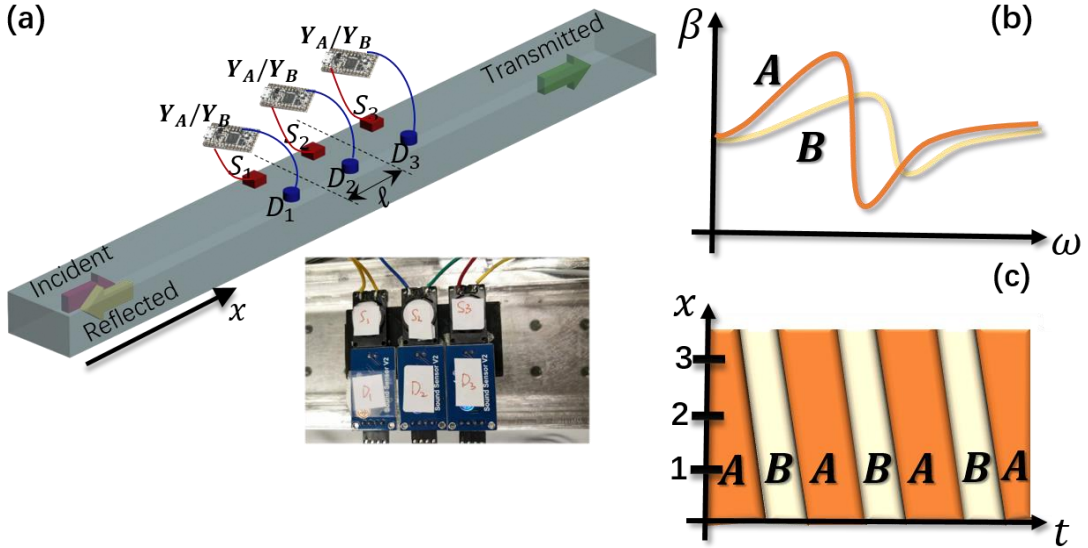

FIG. 1. (a) Schematic diagram of the spacetime metamaterial. Digitally virtualized atoms are assembled on a one-dimensional waveguide with waves propagating along the x-axis. Each atom consists of a microphone and speaker interconnected by an external microcontroller where the response function (convolution kernel) $Y_A$ and $Y_B$ alternate in time. Inset: a photograph of the waveguide with 3 virtualized atoms, where the lattice constant between neighbouring atoms is $\ell = 2$cm. (b) Two static configurations A and B give rise to their corresponding compressibility spectra of the metamaterial. (c) The two configurations are then switched in alternating sequence periodically in time $t$ and three atoms (black horizontal bars) can be synchronized in a way that the modulation "wavefront" propagates at an angle in realizing a spacetime metamaterial.

To begin our discussion, we consider a temporal metamaterial in switching between two static configurations A and B. In this case, we set $a(t)$ as a function in periodically switching between two constants $a_A$ and $a_B$ (or $Y(t,t')$ switching between $Y_A(t')$ and $Y_B(t')$) with duty cycle $\xi:1-\xi$ within a modulation period of $T$. Static configuration A/B means the situation when $a(t)$ is a time-invariant constant $a_A/a_B$. It in turns corresponds to a static compressibility spectrum of the metamaterial given by $\beta_{A/B}(\omega) \cong 1 + 2Y_{A/B}(\omega)/i$ with the assumed monopolar scattering response from each atom. The compressibility is defined with respect to the one of air. $Y_{A/B}(\omega)$ is the Fourier transform of $Y_{A/B}(t')$. Figure 1(b) shows schematically the resonating compressibility $\beta_{A/B}(\omega)$ of the two configurations. Within our scheme, the concurrent use of two time variables ($t$ for the actual time and $t'$ for the convolution integral) in $Y$ allows us to specify the meaning of time-varying spectral response. Certainly, the compressibility can be further made inhomogeneous in the spatial domain (in addition to the forementioned temporal domain). When we keep the same kind of modulation for all the three atoms but with a time lag of the modulation from atom 1 to 2 and from atom 2 to 3, our modulation scheme is also able to construct a spatiotemporal effective medium, depicted in Fig. 1(c), in which the modulation "wavefront" now propagates at an angle, rather than to the right hand side in the case of purely temporal effective medium.

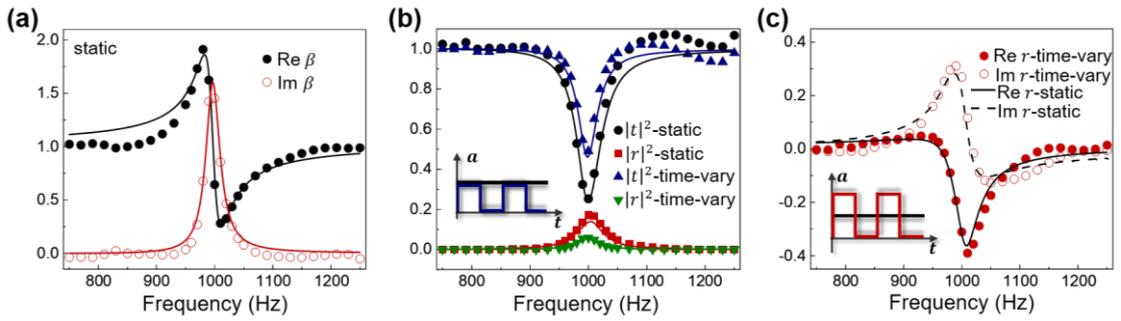

FIG. 2. (a) Compressibility of the static virtualized metamaterial. Symbols show the results extracted from experiment by measuring transmission and reflection coefficients on a single atom, while black and red lines show the instructed Lorentzian resonating model (resonating frequency 1kHz and linewidth 15Hz) in its real and imaginary parts respectively.(b) Transmittance and reflectance for either a static metamaterial of 3 atoms or a time-varying metamaterial with modulation period 140us (7.1kHz) and the 3 atoms are all turned on and off (modulation $a(t)$ switching between 1 and 0, as shown in the inset) every 70us. The symbols show the experimental results while the lines show the theoretical model results predicted by the Lorentzian model of a single atom. (c) The real (solid symbols) and imaginary (empty

symbols) parts of the complex reflection coefficient for the time-varying metamaterial with doubled resonance strength in the first half of the modulation period (modulation $a(t)$ switching between 2 and 0). The black lines are the corresponding theoretical results for an equivalent static metamaterial with $a(t) \equiv 1$ without modulation.

In our current experiment, the modulation period is set to $140\mu s$, being at around 6-10 times smaller than the signal period in our interested frequency range. On the other hand, the lattice constant of the metamaterial is set at 2cm, which is also much smaller than the wavelength of the signal. These two conditions validate the spatiotemporal effective medium description. First, we establish the metamaterial before applying temporal modulation. Figure 2(a) shows the case when we instruct the virtualized metamaterial in generating a static resonating compressibility $\beta(\omega)$. For the ease of extraction, we have only turned on one of the atoms in this stage, the effective medium, with normalized compressibility and density, is then extracted from the complex transmission (t) and reflection (r) coefficients of a single atom [37] by

$$\beta(\omega) = \frac{2ci}{\omega a}\frac{1-r(\omega)-t(\omega)}{1+r(\omega)+t(\omega)}, \qquad (1)$$
$$\rho(\omega) = \frac{2ci}{\omega a}\frac{1+r(\omega)-t(\omega)}{1-r(\omega)+t(\omega)}.$$

The experimentally extracted compressibility corresponds to a Lorentzian resonating model with resonating frequency at $f_0 = 1\text{kHz}$ and resonance linewidth $\gamma = 15\text{Hz}$, shown in the solid and open symbols in Fig. 2(a) for the real and imaginary parts, agreeing to the Lorentzian model in black and red lines respectively. On the other hand, the effective density is very close to the one of air (not shown). In fact, the extracted $\beta$ represents the material property of the metamaterial consisting of a number of atoms as the near-field coupling between neighbouring atoms is negligible. Figure 2(b) shows the measured transmittance and reflectance for a metamaterial constructed by three identical atoms with symbols. They agree well with the theoretical results predicted by the Lorentzian model of compressibility in panel (a) using a transfer-matrix approach in modeling a homogeneous medium with such compressibility and a thickness of 3 unit cells. With the established frequency dispersion of the static metamaterial, we now modulate the effective compressibility in the time domain. In the current case, we choose to turn on the resonance at full-scale in the first half of the modulation period with $a_A = 1$. In the second half of the modulation period, the atom is turned off with $a_B = 0$ and the modulation repeats as time goes on (inset of Fig. 2(b)). As the modulation frequency (7.1kHz) is at least 5 times greater than the signal frequency, it falls into the forementioned temporal effective medium regime, we then expect the signal cannot differentiate the fine feature of the modulation and have to treat the whole medium as an equivalently static effective medium, being called $\beta_{\text{eff}}(\omega)$, which can be theoretically obtained by

$$\beta_{\text{eff}}(\omega) = \xi \beta_A(\omega) + (1 - \xi)\beta_B(\omega), \tag{2}$$

which reduces to $(1 + \beta(\omega))/2$ in the current case (see derivation in Supplementary Materials). As such, we measure again the complex transmission and reflection coefficients of the same metamaterial but under temporal modulation. It is found that the resonance strength is reduced with a shallower transmission dip and reflection peak, as shown in Fig. 2(b). The results predicted by the theoretical $\beta_{\text{eff}}(\omega)$ (using transfer matrix approach) is plotted in solid and dashed blue lines in the same figure, agreeing well with the experimental results. We note that as the nature of the temporal effective medium is to time-averaging the compressibility, we expect the current time-varying metamaterial should be equivalently to a static metamaterial but with doubled resonance strength ($a_A = 2$ and $a_B = 0$), comparing to the current static atom. Such equivalence is shown by comparing the complex reflection coefficient plotted in Fig. 2(c).

Up to now, we have experimentally proved the effective medium is to temporally average the compressibility. In fact, if we compare to the formulation by a frequency non-dispersive but time-varying permittivity and permeability in the case of time-varying electromagnetic effective medium, Ref. [35], it is interesting to find that we should expect temporally averaging bulk modulus (reciprocal of compressibility, or in analogy to reciprocal of permittivity in electromagnetic waves) instead. In our case, the constitutive relationship is provided by the generation of secondary source, being represented in discrete-time signal by the z-transform for each virtualized atom as

$$Y_i^{(d)}(z)S_i(z) = Y_i^{(n)}(z)D_i(z) \qquad (3)$$

where $Y_i(z) = Y_i^{(n)}(z)/Y_i^{(d)}(z)$ is the original specified response function in converting signal at a detector to a speaker at each virtualized atom. In analogy to electric constitutive relationship, the $D_i$ corresponds to the E-field while the $S_i$ corresponds to the polarization field (P). The distribution of the response $Y_i(z)$ into polynomial of $z$ in numerator $Y_i^{(n)}(z)$ and $Y_i^{(d)}(z)$ and whether the time varying part is embedded in the numerator or the denominator depends actually on the implementation algorithm in the digital signal processing program. In our implementation, the change of resonance strength ($a(t)$) is in the numerator $Y_i^{(n)}$ and it in turns translate to temporally averaging $D_i$ and keep $S_i$ constant, giving temporally averaging compressibility $\beta$. On the other hand, if the time-varying part is in the denominator $Y_i^{(d)}$ by having other implementation, the temporally averaging can occur at bulk modulus instead of compressibility.

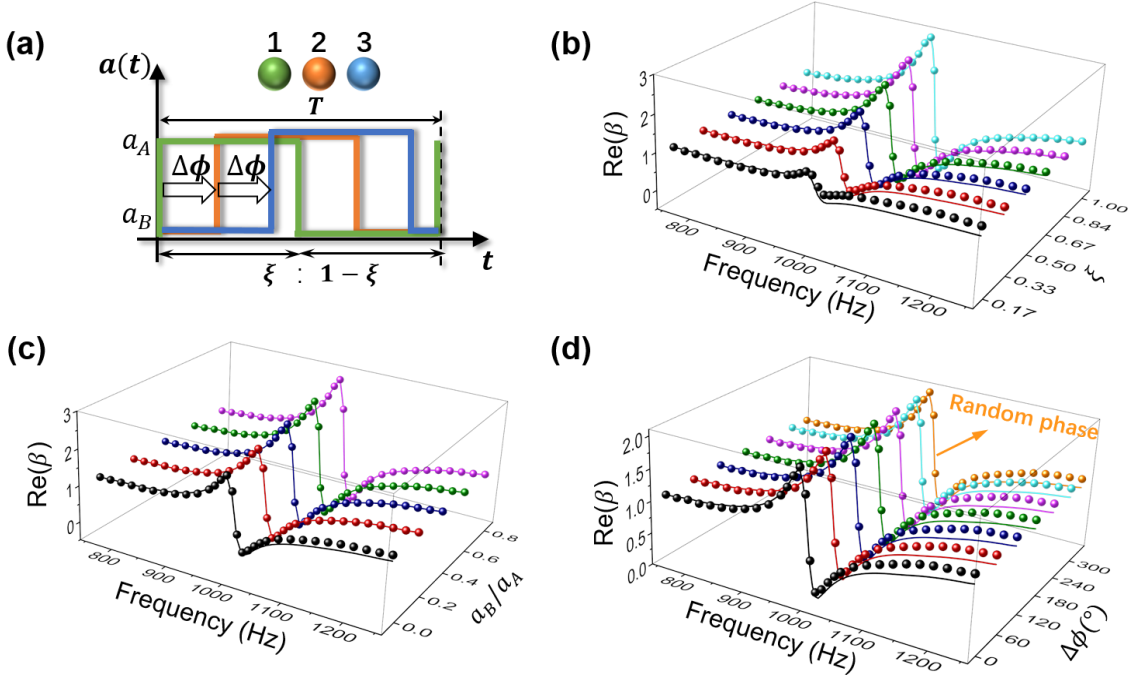

FIG. 3. (a) The analogy of AB-stacking of metamaterial in the temporal domain for 3 atoms (green, orange and blue colors) with dutycycle $\xi$ and modulation phase difference $\Delta\phi$ between neighbouring atoms ($\Delta\phi$ means a shifting of $T\Delta\phi/(2\pi)$ in time). $a_A$ and $a_B$ are the modulation amplitude for two configurations. (b) The real part of measured compressiblity (symbols) of temporal metamaterial of 3 atoms for different duty cycles with fixed $a_A = 1, a_B = 0, \Delta\phi = 0$. The solid lines in corresponding colors show the model results according to the temporal effective medium (i.e. Eq. (2)). (c) Compressiblity for different modulation amplitude $a_B$ (change from 0 to 0.8 with a step of 0.2) with fixed $\xi = 0.5$, $a_A = 1$, $\Delta\phi = 0$. (d) Compressiblity for different $\Delta\phi$ between neighbouring atoms with fixed $\xi = 0.5$, $a_A = 1, a_B = 0$. The brown line represents the case with random modulated phase. The modulation frequency for all cases in (b), (c) and (d) is 4.8kHz.

In order to further explore the flexibility of our time modulation scheme, we study its performance in different situations by varying duty cycle $\xi$, modulation amplitudes ($a_A$ and $a_B$ for two alternating configurations) and modulation phase difference between neighbouring atoms ($\Delta\phi$ between atom 1 and 2 and between atom 2 and 3, which can range from 0 to $2\pi$ in shifting across one whole modulation period), as shown schematically in Fig. 3(a). Figure 3(b) shows the compressibility spectra of temporal metamaterial by changing duty cycle from $\xi = 1/6$ to $\xi = 1$ (corresponding to static one), with compressibility "on" and "off" alternatively with duty cycle $\xi:1-\xi$ in each modulation period. For the duty cycle $\xi = 1/6$, the real part of measured effective compressibility follows a Lorentzian dispersion, as shown with black symbols. As the duty cycle increases from $\xi = 1/6$ (black) to $\xi = 1$ (cyan), the effective compressibility increases proportionally, plotted with symbols with different colors. The solid lines correspond to the model results given by the temporal effective medium Eq. (2), which capture well the experimental results around the resonance frequency with larger signal to noise ratio. Similarly, the larger $\xi$, the larger polarizability, thus model results are more consistent with experimental results. This means that the effective compressibility $\beta_{\text{eff}}(\omega)$ of the temporal metamaterial can be tuned from the one of air to the static result $\beta(\omega)$ at full cycle by changing $\xi$ from 0 to 1. In addition, the modulation strength provides another degree of freedom for tailoring the effective compressibility of temporal metamaterials. For simplicity, we fix $a_A = 1$ and change modulation amplitude $a_B$ from 0 (black) to 0.8 (magenta) in steps of 0.2. Similarly, we extract the effective compressibility (solid dots) and compare them with the model results (solid lines) from Eq. (2), shown in Fig. 3(c). As $a_B/a_A$ increases from 0 to 0.8, the experimental $\beta_{\text{eff}}$ increases proportionally, agreeing well with the model results. From the above discussion, we now know that we have two significant parameters that can be used to flexibly construct the temporal effective compressibility, with the same recipe as multilayer metamaterial in space domain by varying filling fraction and material parameter ratio between the two stacking layers.

For spatial metamaterials, it is well known that the effective material parameters are insensitive when we apply a positional disorder to the metamaterial atoms due to the local resonating nature of the atoms. As an analogy, we would like to ask if the phase difference of time modulation between neighbouring atoms would affect the resonating response or the effective material parameters? To answer this question, we set a phase delay on neighbouring atoms. In Figure 3(d), it is not difficult to find that the effective compressibility of temporal metamaterial is almost insensitive with increasing phase delay $\Delta\phi$ from 0° (black) to 300° (cyan), even setting random modulation phases (brown). The reason of unchanged effective compressibility is that the working frequency is much smaller than the modulation frequency so that the signal arriving each atom cannot distinguish different moments the signal wave hits the atom. As a result, the synchronization between the different atoms in the effective medium regime becomes unimportant, the brown symbols in Fig. 3(d) represent the experimental results for the case when the modulation phase of the different atoms are chosen randomly, agreeing to the theoretical result plotted with assumption of zero modulation phase lag between the different atoms as solid line of the same color.

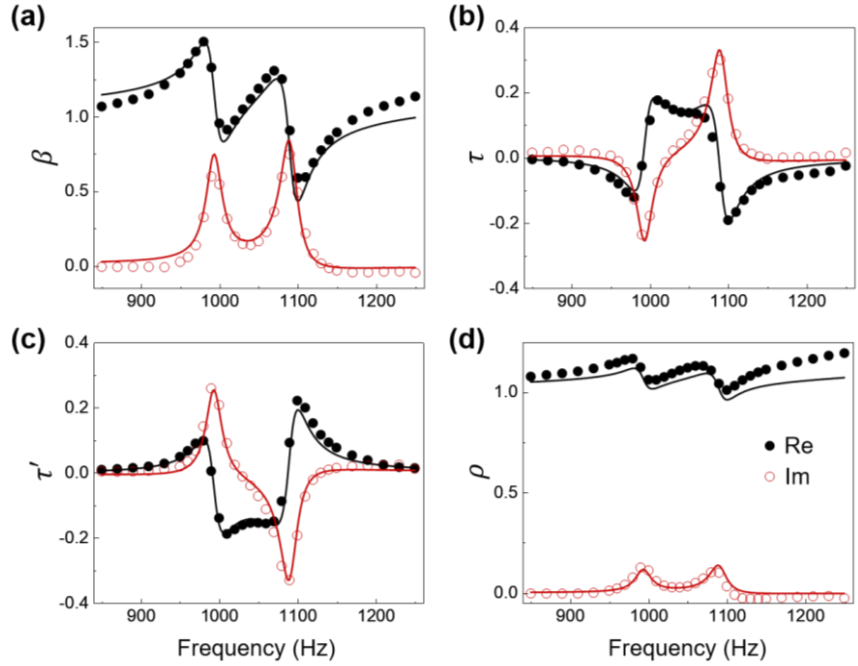

FIG. 4. The extracted constitutive parameters: compressibility $\beta$ (a), density $\rho$ (d) and cross Willis coupling terms $\tau$ (b) and $\tau'$ (c) of a virtual metamaterial consisting of two atoms with different resonance frequency ($f_1 = 1k$Hz and $f_2 = 1.1k$Hz). Here the black (red) symbols represent the real (imaginary) part of experimental results for a time modulated unit-cell, while the solid lines represent the theoretical effective constitutive parameters obtained from temporal effective medium theory.

Finally, we would like to extend our discussion of temporal effective medium to the constitutive parameters other than the compressibility. Specifically, it has been recently established the notion of Willis coupling parameters in both the acoustic and elastic wave domain as an analogy of bianisotropy in electromagnetism [40,41,42]. By applying a temporal modulation on a bianisotropic metamaterial, one can obtain a temporal metamaterial with Willis coupling terms, with all the 4 constitutive matrix elements. Here, we choose to construct a virtual metamaterial consisting of two atoms (atom 1 and atom 2 with separation of 4cm) with different resonating frequencies ($f_1 = 1kHz$ and $f_2 = 1.1kHz$) but the same resonating linewidth $\gamma = 15Hz$ and resonating strength $g = 16Hz$. Consequently, such a structure can generate a monopolar polarizability (symmetric input generates the symmetric scattering field) and a dipolar polarizability (anti-symmetric input generates an anti-symmetric scattering field) at the same time, affecting the bulk modulus and mass density of acoustic metamaterials respectively. On the other hand, the asymmetric structure with resonance can also generate significant cross coupling terms which relate symmetric (anti-symmetric) incident to anti-symmetric (symmetric) scattering components, termed as Willis coupling parameters. Applying a square wave amplitude modulation (modulation frequency $f_m = 7.2kHz$, $a_A = 2, a_B = 0$, and $\xi = 0.5$) on the metamaterial, the generated scattered field satisfies the following constitutive relations [40,43]:

$$\begin{pmatrix} -\varepsilon \\ \mu \end{pmatrix} = \begin{pmatrix} \beta & i\tau \\ i\tau' & \rho \end{pmatrix} \begin{pmatrix} p \\ v \end{pmatrix} \qquad (4)$$

where $\varepsilon$ and $\mu$ represent the volume strain and momentum fields respectively, $p$ and $v$ are the pressure and velocity fields respectively. In our experiment, we can extract the constitutive matrix from the measured scattering matrix (of the temporal metamaterial). In the limit of reciprocal metamaterial, it can be written as

$$\beta(\omega) = \frac{2ci}{3\omega\ell} \frac{(-1 + r_b + r_f - r_b r_f + t^2)}{r_b r_f - (1+t)^2},$$

$$\rho(\omega) = -\frac{2ci}{3\omega\ell} \frac{(1 + r_b + r_f + r_b r_f - t^2)}{r_b r_f - (1+t)^2}, \quad (5)$$

$$\tau(\omega) = -\tau'(\omega) = \frac{2c}{3\omega\ell} \frac{(r_b - r_f)}{r_b r_f - (1+t)^2},$$

where the subscript "f" and "b" are for labeling the forward and backward transmission or reflection coefficients. Figure 4 shows the real and imaginary parts of the four experimentally extracted constitutive parameters: compressibility $\beta$ (in panel (a)), density $\rho$ (in panel (d)) and Willis coupling terms $\tau$, $\tau'$ (in panel (b) and (c)) for the temporal metamaterials with black and red symbols respectively for their real and imaginary parts. Different from the previous cases with only monopolar polarizability (denoted as $\beta$ before), here the temporal metamaterial presents significant dipolar polarizability to give $\rho$ with values deviating from one and cross coupling terms. All constitutive parameters are observed with two resonance frequencies, and $\tau \cong -\tau'$ due to the realized reciprocity of our system. To get the effective medium theory for the temporal Willis metamaterial, we begin from the static constitutive matrices which can be obtained from the atomic responses. Then, we temporally average these constitutive matrices to obtain the final temporal effective medium. Its real and imaginary parts are plotted with black and red lines in the same figure, matching well with the experimentally extracted results in symbols. It is noted that the existence of Willis coupling terms leads to the difference between forward $r_f$ and backward reflection $r_b$ coefficient, while forward transmission and backward transmission is the same ($t = t_f = t_b$) due to the reciprocity. These signature carries over to the temporal Willis metamaterial.

In summary, we have implemented an acoustic metamaterial for realizing spatiotemporal effective medium, in which both the spatial and temporal modulation periods are much smaller than the signal wavelength and period. The established effective medium in our implementation of temporal modulation supports the temporal averaging of the constitutive parameters including compressibility, density and Willis coupling terms. In the spatiotemporal effective medium regime, the modulation phase difference between the different atoms has negligible effect on the effective medium. Our implementation will be useful as a testbed for future exploration of various proposed spatiotemporal phenomena in exploiting temporal modulation-induced non-reciprocity, broadband response, etc. We also note that in addition to the established model on spatiotemporal effective medium, we have also used Floquet band model and time-domain simulation in checking against with the experimental results.

The work is supported by Hong Kong Research Grants Council (RGC) grants (C6013-18G, 16303019).